# Standard Cell Library Design and Optimization Methodology for ASAP7 PDK

(Invited Paper)

Xiaoqing Xu[1], Nishi Shah[1], Andrew Evans[1], Saurabh Sinha[1], Brian Cline[1], and Greg Yeric[1]

[1]ARM Inc., Austin, TX, USA

{xiaoqing.xu,nishi.shah,andrew.evans,saurabh.sinha,brian.cline,greg.yeric}@arm.com

## ABSTRACT

Standard cell libraries are the foundation for the entire back-end design and optimization flow in modern application-specific integrated circuit designs. At $7nm$ technology node and beyond, standard cell library design and optimization is becoming increasingly difficult due to extremely complex design constraints, as described in the ASAP7 process design kit (PDK). Notable complexities include discrete transistor sizing due to FinFETs, complicated design rules from lithography and restrictive layout space from modern standard cell architectures. The design methodology presented in this paper enables efficient and high-quality standard cell library design and optimization with the ASAP7 PDK. The key techniques include exhaustive transistor sizing for cell timing optimization, transistor placement with generalized Euler paths and back-end design prototyping for library-level explorations.

## 1. INTRODUCTION

The state-of-the-art back-end design flow builds upon a presumably well-optimized standard cell (SC) library. However, at the $7nm$ technology node and beyond, such presumption is becoming extremely challenging and expensive due to ever increasing design complexities, such as discrete transistor sizing from FinFETs [1] and restrictive design rules from lithography [2]. Modern SC library design requires a specific process design kit (PDK) to produce logic and sequential elements with various driving strengths.

A single SC design and optimization typically involves two phases, i.e., schematic and layout design. The introduction of FinFET limits the flexibilities of a schematic design when translating a logic/sequential element into a transistor-level netlist [1]. Moreover, geometric scaling results in significant challenges in terms of SC layout design, especially when accommodating restrictive design rules from lithography. Specifically, the SC layout design is restricted within a fixed height, i.e., SC architecture, which incurs large turnaround time for transistor-level placement and routing in a human-driven procedure. This further leads to a wide research domain of design technology co-optimization (DTCO) [3, 4], which is beyond the scope of SC library design and optimization methodology in this study.

Despite the renowned challenges and difficulties aforementioned, few comprehensive studies have been performed for SC design and optimization methodology at the $7nm$ technology node and beyond. NanGate proposes a 12-track height SC design at the $15nm$ node, where the underlying FinFET geometries and parasitic extractions are not fully revealed [5]. This means the NanGate 15nm PDK [6] provides very limited value for academic researchers to perform SC design and optimization studies. Recent DTCO studies focus on a small set of SCs and basic block-level placement and routing studies to explore the impact of lithography design rules and SC architectures [2, 7–9]. However, very few SC design and optimization techniques are discussed for complex logic/sequential cells. More importantly, none of them are publicly available, which prevents academic researchers from exploring novel SC design and optimization techniques.

Therefore, this study proposes a set of design and optimization techniques to create publicly-available SC libraries with ASAP7 PDK [10–12]. We first discuss exhaustive transistor sizing for cell timing optimization taking advantage of the discrete transistor sizing for a FinFET-based SC design. Generalized Euler path theory [13] is adopted to generate high-quality transistor placement results. The generalized Euler paths lead to a much larger solution space than that of conventional Euler path theory, while accommodating pin accessibility, pin capacitance, diffusion breaks and dummy gate insertion. We further explore SC architecture options (9-track and 7.5-track) and SC library richness by embedding various SC libraries into a full back-end design flow. The library-level optimizations are based on the final power, performance and area metrics from an entire back-end design flow, which also generates valuable baseline results for future research studies. Therefore, we anticipate a much broader academic impact, while leveraging very recent research on SC library evaluation and optimization [14–16], automatic SC synthesis [17–19], SC-driven placement [20–24] and routing [25–28] towards a full synthesis research flow [29].

The rest of this paper is organized as follows: Section 2 discusses the SC architecture options, transistor sizing and generalized Euler paths for transistor placement. Section 3 embeds various SC libraries into an entire back-end design flow to explore different SC architecture options and library richness. Section 4 concludes the paper.

## 2. STANDARD CELL LIBRARY DESIGN

### 2.1 Standard Cell Architecture

The state-of-the-art SC layout design follows a specific SC architecture, which requires a fixed SC height in terms of horizontal metal (metal-2) routing tracks. This is consistent with a horizontal row-based structure for power/ground rails and SC placement in the back-end design flow. Fig. 1 illustrates the basic SC architecture options, i.e., 9-track and 7.5-track, which are compatible with the technology parameters in the ASAP7 PDK as shown in Table 1. For the 9-track architecture in Fig. 1(a), the SC height is 9 metal-2 pitches, which allows 8/9 metal-1 and metal-2 tracks for signal routing. For the 7.5-track architecture in Fig. 1(b), the SC height is 7.5 metal-2 pitches, which allows 5.5 metal-1 tracks and 6 metal-2 tracks for signal routing. The track resources on the top and bottom of the cell are typically reserved for power/ground rails. Metal-1 tracks are primarily used for intra-cell routing, while occasional metal-2 usage is allowed but discouraged in order to provide the router as much flexibility as possible with metal-2 routing.

**Table 1:** SC architecture related parameters

| Fin pitch (nm) | 27 | |
|---|---|---|
| metal-1/metal-2 pitch (nm) | 36 | |
| SC architecture | 9-track | 7.5-track |
| total # of fins | 12 | 10 |
| # of fins for transistor | 4 | 3 |
| # of metal-1 tracks for signal routing | 8 | 5.5 |
| # of metal-2 tracks for signal routing | 8 | 6 |
| metal-2 and metal-1 track offset (nm) | 0 | 9 |

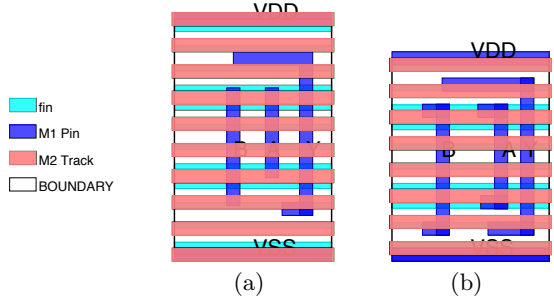

**Figure 1:** SC architecture, (a) 9-track, (b) 7.5-track.

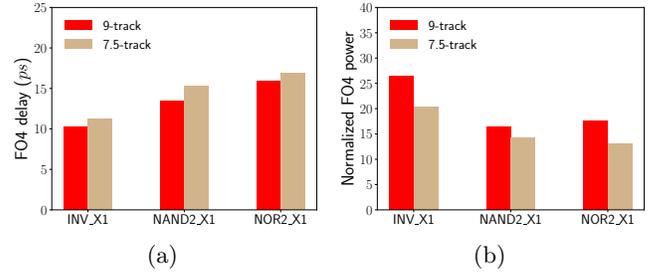

**Figure 2:** FO4 circuit comparisons, (a) delay, (b) power.

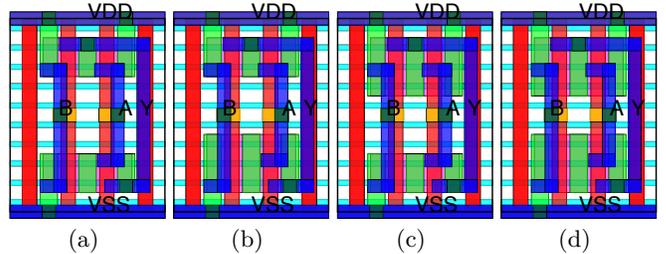

**Figure 3:** Exhaustive transistor sizing for NAND2_X1N with 7.5-track architecture, (a) (2p, 2n), (b) (2p, 3n), (c) (3p, 2n), (d) (3p, 3n).

Although 7.5-track generates more aggressive area scaling compared to the 9-track architecture, it accommodates less number of fins for each transistor, thereby less amount of maximum driving current per transistor. The 7.5-track architecture also leads to more intra-cell routability issues due to less metal-1 track resources. This makes metal-2 usages inevitable for complex cells under the 7.5-track architecture. Moreover, the metal-1 and metal-2 track offset makes the layout more prone to complex design rule violations. When making an SC architecture decision, it is important to explore power, performance and area (PPA) trade-offs at both the SC level and block level. This study explores the SC-level trade-off by comparing the PPA for basic fan-out-4 (FO4) circuits as shown in Fig. 2. Fig. 2(a) shows that, for basic logic cells (INV_X1, NAND2_X1 and NOR2_X1), the 7.5-track SCs generates larger delay than 9-track cells. The reason is that the 7.5-track architecture allows up to 3 fins per transistor, which delivers smaller on-state currents than the 9-track architecture (up to 4 fins per transistor). Meanwhile, the superior SC performance of the 9-track architecture is not free - it comes at the cost of slightly higher power consumptions as shown in Fig. 2(b). However, simple figure-of-merit circuits like SC FO4's do not always tell the full DTCO story [30], so block-level comparisons are also performed by using the various SC libraries to implement basic back-end design flows (see Section 3).

## 2.2 Transistor Sizing

For a given SC architecture in Table 1, the effective gate width of one transistor is limited to a discrete number of fins, i.e., discrete sizing solution space. The single-fin driving currents for PMOS and NMOS are different in ASAP7 PDK [10]. It becomes difficult, even impossible to fully balance the rising and falling edges for a logic gate. This is because FinFET loses the continuous sizing flexibility compared to the SC-level sizing for planar transistors.

However, the discrete sizing solution space enables exhaustive transistor sizing to decide the best number of fins for each PMOS or NMOS transistor within a basic logic SC. Fig. 3 demonstrates the exhaustive transistor sizing for the minimum-sized 2-input NAND gate (referred as NAND2_X1 for the remainder of the paper). We set the minimum and maximum number of fins as 2 and 3, respectively, for each transistor. The neighboring transistor sharing source/drain regions shall have the same number of fins due to restrictive design rules for the active layer. Fig. 3(a) shows a combination of 2 fins for PMOS and NMOS, denoted as (2p, 2n). Other combinations are exhaustively enumerated in Fig. 3(b), 3(c) and 3(d). Then, parasitic extractions and SPICE simulations are performed to collect the rising and falling delay/slew for each combination. The final transistor sizing for a basic logic SC is decided based on the most balanced rising and falling edges from simulation results. Thus, for a set of basic logic SCs, we obtain a set of most balanced pull-up networks (PMOS's) and pull-down networks (NMOS's). For complex logic and sequential SCs, the sizing is further done based on the most balanced pull-up and pull-down networks from basic logic SCs, where human-driven cell tuning is needed depending on each SC.

## 2.3 Transistor Placement with Euler Path

For an SC design after transistor sizing, i.e., schematic design, the layout design is to perform transistor placement and routing under a fixed SC architecture. Euler path theory [31] is a widely adopted optimization technique for transistor placement, which generates optimal source/drain sharing for compact SC area.

Fig. 4 demonstrates the multi-stage and-or-inverter cell (AOI31_X2) layout design with the consistent Euler paths in Eulerian graphs. For the schematic in Fig. 4(a), the undirected multigraph for PMOS (pull-up logic) and NMOS (pull-down logic) are shown in Fig. 4(b). For a multigraph for PMOS or NMOS, each transistor and connection is denoted with a graph edge and node, respectively. A common Euler path, i.e., (A0,B0,B0,A0,A1,A2,A2,A1), can be consistently found in the transistor connection graphs for pull-up and pull-down logic because the multigraphs for NMOS/PMOS are Eulerian. This Euler path further translates into the area-optimal layout shown in Fig. 4(c). It is area optimal because it is impossible to achieve another layout design with an area smaller than that in Fig. 4(c) for AOI31_X2 [31]. It shall be noted that there may exist multiple area-optimal Euler paths for a logic SC. We break the tie by favoring better pin accessibility and smaller pin capacitance. For example, Euler path (B0,A0,A1,A2,A2,A1,A0,B0) also leads to area-optimal layout for AOI31_X2 in Fig. 4(a), but it induces worse pin accessibility. The pin capacitance is also worse because of much larger metal areas from the pin shapes.

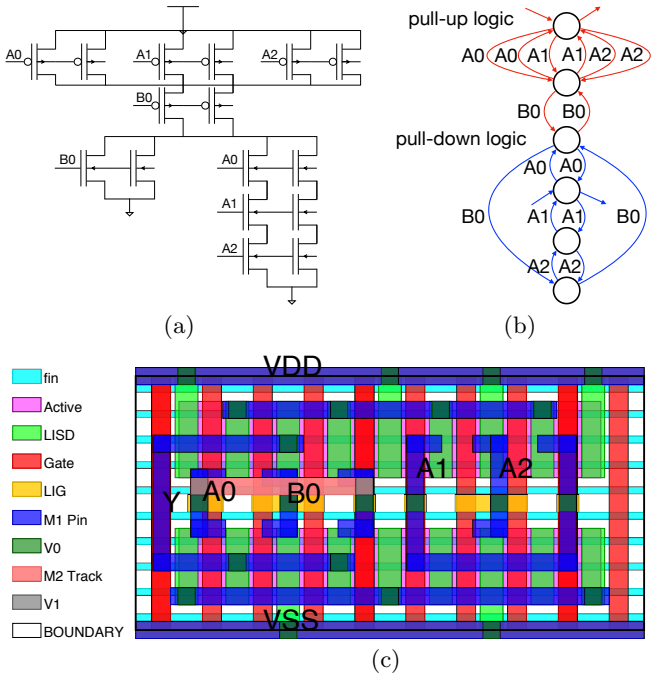

**Figure 4:** Euler path and layout for AOI31_X2, (a) circuit schematic, (b) a consistent Euler path (A0,B0,B0,A0,A1,A2,A2,A1) for pull-up and pull-down logic, (c) area-optimal layout.

For complex logic and sequential SCs, the corresponding multigraphs are not guaranteed to be Eulerian. Moreover, it becomes difficult to directly apply the Euler path theory and generate area-optimal layout under restrictive design rules, especially when we also need to obtain better pin accessibility, pin capacitance and limited metal-2 routing usages. Therefore, we generalize the Euler paths to the cases of diffusion breaks and dummy gate insertions [13]. An example of layout design for transmission-gate-based 2-input multiplexer (MXT2_X1) is shown in Fig. 5. The pull-up and pull-down logics are not fully complementary logics due to the insertion of pass gates for fewer transistor counts. Moreover, it is not possible to find a layout with source/drain regions fully shared due to the connections between source/drain regions and transistor gates, such as net "ns0" and "ny" in Fig. 5(a).

A consistent Euler path, i.e., (S0,A,S0,ns0,B,0,ny), among pull-up and pull-down logic, is shown in Fig. 5(b), where "0" represents the diffusion break in the layout [32]. However, input pin A is completely blocked for pin access, which makes this area-compact layout solution infeasible for practical designs. Inconsistent Euler paths, i.e., (S0,ny,0,S0,ns0,0,B,A) for pull-up logic and (S0,ny,0,0,ns0,S0,B,A) for pull-down logic, are shown in Fig. 5(c). This further generates a pin-accessible layout design, where "0" denotes dummy gate insertion or diffusion break.

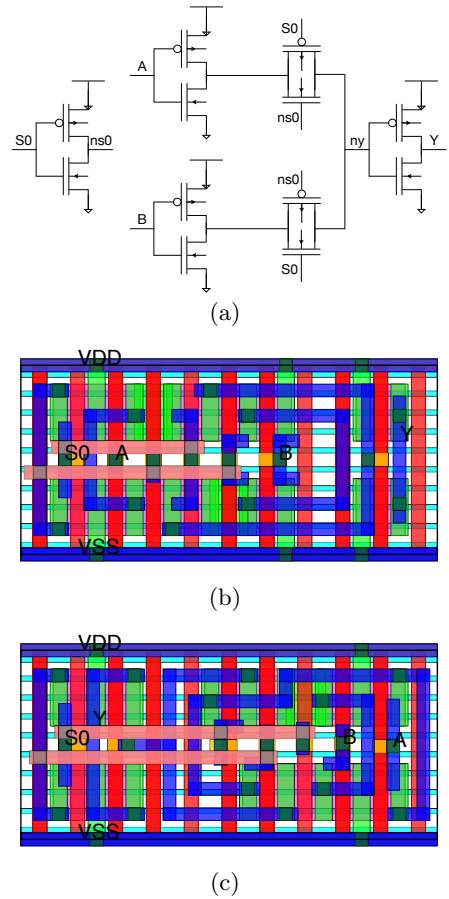

**Figure 5:** MXT2_X1 layout with different generalized Euler path, (a) circuit schematic, (b) for Euler path: pull-up and pull-down (S0,A,S0,ns0,B,0,0,ny), input pin A is blocked for pin access, (b) for Euler path: pull-up (S0,ny,0,S0,ns0,0,B,A), pull-down (S0,ny,0,0,ns0,S0,B,A), the transistor placement leads to better pin access.

This study adopts generalized Euler paths to accommodate dummy gate insertion and diffusion break, which essentially explores a sufficiently large solution space for high-quality transistor placement and routing results. For complex SCs, this further leads to superior layout solutions in terms of intra-cell routability and pin accessibility as shown in Fig. 5.

## 3. DESIGN SYNTHESIS AND EXPLORATION

### 3.1 Synthesis Flow

The SC library design only yields a set of cells characterized and optimized in isolation. A well-optimized SC library should further deliver high-quality results in terms of PPA from a full synthesis flow. This study adopts Arm® Cortex®-M0 processor from Arm DesignStart™ portal [33] to enable the design synthesis and exploration. The basic synthesis w/ placement and routing flow is demonstrated in Fig. 6. Given an SC library, it first translates the design RTL into the gate-level netlist with the logic synthesis engine from Cadence [Genus Synthesis Solution, v15.12] [34]. The partitioning & floorplanning and placement & routing are performed with Cadence Innovus [Innovus(TM) Implementation System, v15.10] [35] to achieve the design closure, which further generates quality metrics in terms of PPA. We empirically set the clock period (CP) constraints for the logic synthesis shown in Table 2. For high-frequency design targets, such as 1 GHz, the clock periods for logic synthesis are much smaller than those from the target frequency. This is to enable better timing closure in the single-pass flow shown in Fig. 6.

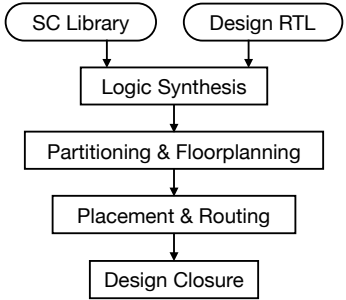

**Figure 6:** SC library-based flow for synthesis w/ placement and routing.

**Table 2:** Synthesis flow settings

| Target frequency (GHz) | 1.0 | 0.9 | 0.8 | 0.7 | 0.6 | 0.5 |
|---|---|---|---|---|---|---|
| Logic synthesis CP (ps) | 850 | 850 | 900 | 1100 | 1400 | 2000 |

We design and optimize 7.5-track and 9-track SC libraries independently, in terms of transistor sizing and layout optimization. We deliver alpha-version 7.5-track and 9-track SC libraries, which include a rich family of SC types and driving strengths. To demonstrate the importance of library richness, we further compose the minimum library, which is a subset of basic logic and sequential SCs from its corresponding alpha library. In a minimum library, we have excluded complex logic gates, such as multiplexers, and SCs with high driving strengths. We use each library within the synthesis w/ placement and routing flow to generate PPA results for the Cortex-M0 processor, which leads to the comparisons under different SC architecture and library richness in Table 3. Specifically, an alpha-version and minimum library are denoted as "alpha lib" and "min lib", respectively. SC architecture is denoted as "SC arch.". The performance is quantified with design target frequency ("Freq."), worst negative slack ("WNS") and total negative slack ("TNS"). The fast and slow corners are used for hold and setup timing analysis, respectively. A negative value of "WNS" means timing failures, while a positive value denotes timing closure. The power is evaluated with total power consumption ("Power") and leakage power ("Leakage"). The activity factor as 20% for power calculations with Innovus at typical corner. The area is quantified with placement utilization ("Util"), total gate count ("Gate #") and total gate area ("Area").

## 3.2 Explore Standard Cell Architecture

The SC library design and optimization can only deliver SC-level PPA comparisons for different SC architecture options. This means only a very bad SC architecture with inferior PPA among all types of SCs can be eliminated. However, it does not happen among typical SC architecture options, i.e., 9-track and 7.5-track, for the ASAP7 PDK. This is because 9-track and 7.5-track architecture represent different

**Table 3:** Synthesis (w/ placement and routing) result comparisons under different SC architecture and library richness

| | SC arch. | Freq. (GHz) | Power (mW) | Leakage ($\mu W$) | WNS (ps) | TNS (ps) | Util. (%) | Gate # | Area ($\mu m^2$) |
|---|---|---|---|---|---|---|---|---|---|
| min lib | 7.5-track | 1.0 | 2.10 | 1.69 | -41 | -10385 | 69.5 | 13956 | 1491.5 |
| | | 0.9 | 1.77 | 1.51 | -18 | -354 | 65.2 | 13248 | 1398.3 |
| | | 0.8 | 1.48 | 1.42 | 0 | 0 | 63.6 | 13302 | 1360.2 |
| | | 0.7 | 1.28 | 1.38 | 39 | 0 | 68.8 | 11956 | 1324.8 |
| | | 0.6 | 1.11 | 1.29 | 46 | 0 | 70.4 | 11719 | 1276.7 |
| | | 0.5 | 0.87 | 1.23 | 48 | 0 | 71.1 | 11430 | 1239.8 |
| | 9-track | 1.0 | 2.38 | 2.29 | -92 | -24214 | 64.4 | 16489 | 1936.2 |
| | | 0.9 | 1.87 | 1.90 | 2 | 0 | 61.6 | 15071 | 1734.2 |
| | | 0.8 | 1.92 | 2.24 | -2 | -2 | 67.7 | 15371 | 1898.1 |
| | | 0.7 | 1.47 | 1.72 | 19 | 0 | 69.6 | 12049 | 1589.1 |
| | | 0.6 | 1.41 | 1.75 | 1 | 0 | 70.1 | 12412 | 1603.2 |
| | | 0.5 | 1.21 | 1.77 | 6 | 0 | 71.0 | 12709 | 1618.2 |
| alpha lib | 7.5-track | 1.0 | 2.26 | 1.90 | -32 | -893 | 66.4 | 13454 | 1537.9 |
| | | 0.9 | 2.02 | 1.88 | -14 | -282 | 63.7 | 13351 | 1518.0 |
| | | 0.8 | 1.53 | 1.65 | 8 | 0 | 68.6 | 11631 | 1409.7 |
| | | 0.7 | 1.29 | 1.46 | 9 | 0 | 69.8 | 10832 | 1306.9 |
| | | 0.6 | 1.00 | 1.19 | 50 | 0 | 70.4 | 9827 | 1177.9 |
| | | 0.5 | 0.74 | 1.10 | 37 | 0 | 70.9 | 9305 | 1133.2 |
| | 9-track | 1.0 | 2.21 | 2.03 | 0 | 0 | 65.7 | 11832 | 1646.9 |
| | | 0.9 | 2.05 | 2.18 | 0 | 0 | 68.7 | 11824 | 1715.8 |
| | | 0.8 | 1.73 | 1.97 | 45 | 0 | 68.8 | 10961 | 1622.8 |
| | | 0.7 | 1.41 | 1.63 | 55 | 0 | 69.9 | 9931 | 1463.5 |
| | | 0.6 | 1.14 | 1.50 | 59 | 0 | 70.6 | 9558 | 1396.1 |
| | | 0.5 | 0.90 | 1.42 | 2 | 0 | 70.9 | 9324 | 1360.7 |

PPA trade-offs at the SC level. As shown in Table 3, the 9-track libraries (min lib or alpha lib) lead to higher power consumption and gate area than the 7.5-track libraries. However, 9-track libraries can push the design closure to a higher frequency. Specifically, the 9-track libraries can obtain timing closure at 0.9 GHz and 1.0 GHz for min lib and alpha lib, respectively, while 7.5-track libraries lead to timing failures in both cases. For high-frequency designs, the advantages of 9-track libraries over 7.5-track libraries are further demonstrated in Fig. 7 (min lib) and Fig. 8 (alpha lib). Fig. 7(a) and Fig. 7(b) illustrates the WNS and TNS after the detailed routing phase across a range of target frequency under minimum libraries. The 7.5-track min lib generates negative values for "WNS" and "TNS", i.e., timing failures, when reaching the target frequency of 0.9 GHz. Similar observations can be made for alpha libraries in Fig. 8.

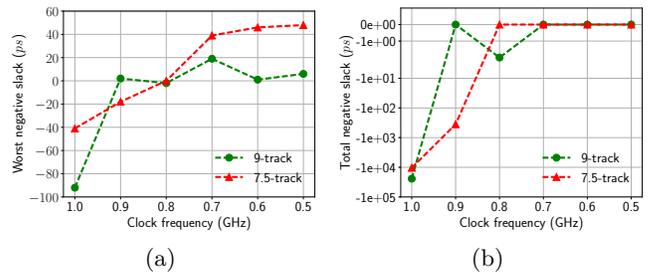

**Figure 7:** Timing results from synthesis (w/ placement and routing) with 9-track and 7.5-track minimum library, (a) WNS, (c) TNS.

## 3.3 Explore Library Richness

The library richness plays an important role in the design closure, because a richer family of SCs results in much larger solution space and design flexibility for a synthesis flow. In this study, the alpha lib includes complex multiplexers and high driving strengths SCs that do not exist in the min lib. For timing-closed designs, the alpha lib generates much better PPA than the min lib as shown in Table 3. This observation holds for most of the design synthesis results. However, for 7.5-track architecture, the "Power" and "Area" from min lib is slightly smaller, i.e., $0.05mW$ and $49.5\mu m^2$, respectively, than those from alpha lib at 0.8 GHz . The reason is that, in this case, the alpha lib uses more complex logic SCs in the final synthesized netlist to close timing. Those complex logic SCs consume larger combinational power to generate better timing results than the synthesized netlist from min lib, i.e., 8 $ps$ vs 0 $ps$ for the "WNS".

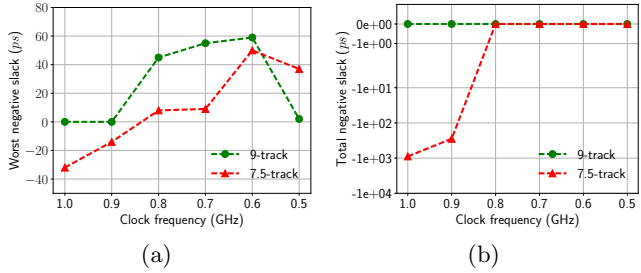

**Figure 8:** Timing results from synthesis (w/ placement and routing) with 9-track and 7.5-track alpha library, (a) WNS, (b) TNS.

Fig. 9 further demonstrates the importance of library richness when pushing the design closure to a higher frequency. Fig. 9(a) and Fig. 9(b) compare the "WNS" and "TNS", respectively, between min lib and alpha lib under the 9-track architecture. At 0.9 GHz and 1.0 GHz, the alpha lib can close the timing while the min lib leads to timing failures, i.e., negative values for "WNS" and "TNS". For other low-frequency cases, the alpha lib consistently obtains better "TNS" than the min lib, which represents much larger flexibility for further power and area optimizations. As shown in Fig. 10, there may exist exceptions for 7.5-track libraries in terms of the observations aforementioned. However, the alpha lib still generates better power and area results than the min lib for most of the cases as shown in Table 3.

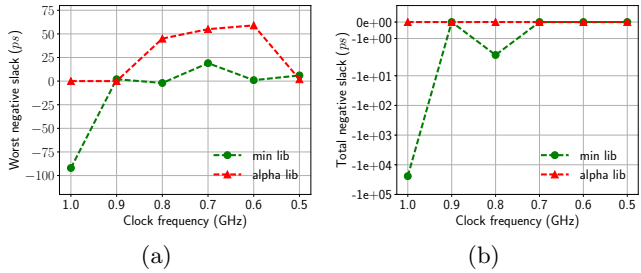

**Figure 9:** Timing results from synthesis (w/ placement and routing) with minimum library (min lib) and alpha library (alpha lib) under 9-track architecture, (a) WNS, (c) TNS.

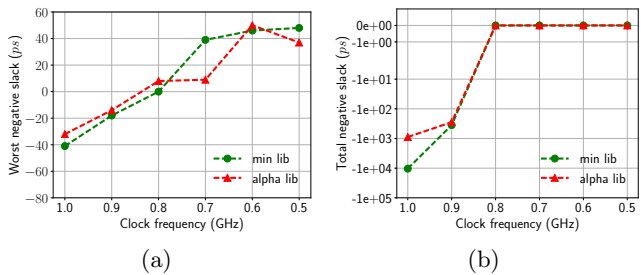

**Figure 10:** Timing results from synthesis (w/ placement and routing) with minimum library (min lib) and alpha library (alpha lib) under 7.5-track architecture, (a) WNS, (b) TNS.

For SC library richness, we further anticipate future research from the following direction. The industrial routine for SC library design and optimization still follows a human-driven procedure. It becomes extremely tedious if an SC library consists of an arbitrarily large set of logic and sequential elements. To avoid huge turnaround time from significant manual design efforts, it becomes pivotal to analyze the importance of each SC type at each driving strength. For example, it is possible that a complex logic SC seldom or even never gets selected by a logic synthesis engine. Then, incorporating many such SCs into the SC library is less meaningful, compared to other frequently used SCs. Although this is a chicken-and-egg issue for one technology, such library richness exploration delivers valuable design experiences for future development efforts.

## 4. CONCLUSION

This paper proposes a set of SC library design and optimization techniques with the ASAP7 PDK. Exhaustive transistor sizing is deployed to balance the falling and rising edges of SC operations while taking advantage of limited sizing flexibilities from FinFET transistors. Generalized Euler paths are adopted to enable a sufficiently large transistor placement solution space, which becomes inevitable for high-quality SC layout in $7nm$ technology node and beyond. We further perform design explorations to enable library-level optimizations based on final PPA metrics from an entire synthesis flow, which leads to valuable baseline results for future academic studies.